\documentclass{aa}
\usepackage{graphicx}
\usepackage{txfonts}
\usepackage{natbib}
\usepackage{ulem}
\bibpunct{(}{)}{;}{a}{}{,} 

\begin{document}

   \title{VLBI detection of the HST-1 feature in the M\,87 jet at 2\,cm}


   \author{C. S. Chang\inst{1}
          \and
          E. Ros\inst{1,2}
          \and
          Y. Y. Kovalev\inst{3,1}
          \and
          M. L. Lister\inst{4}
          }

   \institute{ Max-Planck-Institut f\"ur Radioastronomie, Auf dem H\"ugel 69, D-53121 Bonn, Germany \\
              \email{cschang@mpifr.de, ros@mpifr.de}
         \and
              Departament d'Astronomia i Astrof\'{\i}sica, Universitat de Val\`encia, E-46100 Burjassot, Spain
         \and
             Astro Space Centre of Lebedev Physical Institute, Profsoyuznaya 84/32, 117997 Moscow, Russia\\
             \email{yyk@asc.rssi.ru}
	\and
             Department of Physics, Purdue University, 525 Northwestern Avenue, West Lafayette, IN 47907, USA\\
             \email{mlister@physics.purdue.edu}
             }

   \date{Submitted; \today}


  \abstract
{A bright feature 80\,pc away from the core in the 
powerful jet of M\,87 shows highly unusual properties. Earlier radio, 
optical and X-ray observations have shown that this feature, 
labeled HST-1, is superluminal, and is possibly connected with the 
TeV flare detected by HESS in 2005. It has been claimed that this feature might have a blazar nature, due to these properties. } 
{To examine the possible 
blazar-like nature of HST-1, we analyzed $\lambda$2\,cm VLBA archival data 
from dedicated full-track observations and the 2\,cm survey/MOJAVE 
VLBI monitoring programs obtained between 2000 and 2009. } 
{Applying VLBI wide-field imaging techniques, the HST-1 region was imaged at 
milliarcsecond resolution. } 
{Here we present the first 2\,cm 
VLBI detection of this feature in observations from early 2003 to early 2007, and analyze its evolution over this 
time.  Using the detections of HST-1, we find that the projected apparent speed is 0.61$\pm$0.31$c$. A comparison of the VLA and VLBA flux densities of this feature indicate that is mostly resolved on molliarcsecond scales. This feature is optically thin ($\alpha \sim -0.8$ for S\,$\propto\nu^{+\alpha}$) between $\lambda$2\,cm and $\lambda$20\,cm.} 
{We do not find evidence of a blazar nature for HST-1.}

  \keywords{Radio continuum: galaxies - Techniques: high angular resolution  - Techniques: interferometric - Galaxies: active - Galaxies: jets}

  \maketitle


\section{Introduction \label{sec:intro}}

Active Galactic Nuclei (AGN) are among the most 
energetic phenomena in the Universe, and they have 
been heavily studied since their initial discovery by \cite{seyfert43}. 
Although there are many clues that imply that a super-massive 
black hole (SMBH) is the engine launching the powerful 
jet \citep{urry95}, the exact mechanism remains unknown.


\begin{table*}[t!]
\centering
\caption{Journal of VLBA $\lambda$2\,cm observations of M\,87$^\mathrm{a}$ .    }
\begin{tabular*}{1.0\textwidth}{@{}p{1cm}p{1cm}cp{0.3cm}crp{0.9cm}p{0.9cm}rrrcc@{}}
\hline
\hline
      & Exp.\ & $t_\mathrm{int}$$^\mathrm{b}$ & N$_{\mathrm{ant}}$  & \multicolumn{2}{c}{Beam$^\mathrm{c}$} & S$^{\mathrm{VLBA,M87}}_\mathrm{total}$ & S$^{\mathrm{M87}}_\mathrm{peak,A}$ & S$^{\mathrm{HST-1}}_\mathrm{total}$ & S$^\mathrm{HST-1}_\mathrm{peak,A}$ & S$^\mathrm{HST-1}_\mathrm{peak,B}$  & rms$^\mathrm{d}_\mathrm{A}$ & rms$^\mathrm{d}_\mathrm{B}$ \\

Epoch & Code  & \scriptsize{[min]} & & \scriptsize{size[mas]} &$\phi$\scriptsize{[degree]} & \scriptsize{[Jy]} & \scriptsize{[Jy\,beam$^{-1}$]} &\scriptsize{[mJy]} &\scriptsize{[mJy\,beam$^{-1}$]} & \scriptsize{[mJy\,beam$^{-1}$]} & \scriptsize{[$\mu$Jy\,beam$^{-1}$]} & \scriptsize{[$\mu$Jy\,beam$^{-1}$]}  \\
\hline
2000.06 & \texttt{BK073A}$^\mathrm{e,f}$  & 476&11 & 2.07$\times$1.35 & $-7$  & 2.26 & 1.24 &$<$1.36  &$<$0.26 &$<$0.23 & 79 & 68 \\
2000.35 & \texttt{BK073B}$^\mathrm{e,f}$  & 476&11 & 1.88$\times$1.31 & $-13$ & 2.33 & 1.29 &$<$1.42  &$<$0.20 &$<$0.21 & 70 & 71 \\
2000.99 & \texttt{BK073C}$^\mathrm{e,f}$  & 476&10$^\mathrm{j}$ & 2.12$\times$1.52 & $-3$  & 2.46 & 1.33 &$<$1.84  &$<$0.25 &$<$0.33 & 100 & 92 \\
2001.99 & \texttt{BR077D}$^\mathrm{g,h}$  & 68 &10 & 1.74$\times$1.09 & $-8$  & 2.74 & 1.42 &$<$5.24  &$<$0.66 &$<$0.70 & 212 & 262 \\
2002.25 & \texttt{BR077J}$^\mathrm{g,h}$  & 57 &10 & 1.77$\times$1.13 & $-6$  & 2.54 & 1.40 &$<$5.84  &$<$0.80 &$<$0.81 & 300 & 292 \\
2003.09 & \texttt{BL111E}$^\mathrm{g}$    & 54 &10 & 1.85$\times$1.30 & $-8$  & 2.85 & 1.57 &   3.98  & 1.01   &   2.13 & 218 & 206 \\
2004.61 & \texttt{BL111N}$^\mathrm{g}$    & 63 &10 & 1.88$\times$1.16 & $-11$ & 2.32 & 1.27 &   22.03 & 4.14   &   8.76 & 198 & 210 \\
2004.92 & \texttt{BL111Q}$^\mathrm{g}$    & 64 &10 & 1.95$\times$1.23 & $-14$ & 2.41 & 1.39 &   23.59 & 2.70   &   9.97 & 223 & 295 \\
2005.30 & \texttt{BL123E}$^\mathrm{g}$    & 63 & 9$^\mathrm{k}$ & 2.06$\times$1.37 & $-16$ & 2.35 & 1.37 &   19.72 & 2.57 & 7.03 & 237 & 226\\
2005.85 & \texttt{BL123P}$^\mathrm{g}$    & 63 &10 & 2.11$\times$1.25 & $-17$ & 2.39 & 1.31 &   19.93 & 3.67   &   9.53 & 188 & 208 \\
2006.45 & \texttt{BL137F}$^\mathrm{g}$    & 22 &10 & 1.85$\times$1.22 & $-10$ & 2.51 & 1.51 &$<$6.96  &$<$1.02 &$<$1.44 & 295 & 348 \\
2007.10 & \texttt{BL137N}$^\mathrm{g}$    & 42 &10 & 1.95$\times$1.23 & $-13$ & 2.67 & 1.53 &   10.49 &   1.14 &   2.94 & 201 & 255 \\
2007.42 & \texttt{BL149AA}$^\mathrm{g}$   & 45 &10 & 1.78$\times$1.25 & $-8$  & 2.69 & 1.46 &$<$7.70  &$<$0.81 &$<$1.13 & 272 & 385 \\
2008.33 & \texttt{BL149AO}$^\mathrm{e}$   & 41 & 9$^\mathrm{j}$ & 1.77$\times$1.10 & $-7$  & 3.24 & 1.79 &$<$3.22  &$<$0.54 &$<$0.75 & 193 & 161 \\
2009.10 & \texttt{BL149BG}$^\mathrm{i}$   & 50 & 9$^\mathrm{l}$ & 2.25$\times$1.52 & $-12$ & 1.99 & 1.13 &$<$5.26 & $<$0.36 &$<$ 0.49  & 223 & 263 \\
\hline
 & & & & & & & & & & \\
\multicolumn{13}{@{}l@{}}{\footnotesize{$^\mathrm{a}$ Observations are in dual-polarization mode, except as indicated }} \\
\multicolumn{13}{@{}l@{}}{\footnotesize{$^\mathrm{b}$ Total scheduled VLBI on-source time}} \\
\multicolumn{13}{@{}l@{}}{\footnotesize{$^\mathrm{c}$ Beam A: natural weighting and tapering of Gaussian full-width half maximum 0.3 at 200\,M$\lambda$}} \\
\multicolumn{13}{@{}l@{}}{\footnotesize{$^\mathrm{d}$ Root-mean-square image noise using natural weighting and tapering with a Gaussian factor of 0.3 at a radius of 200\,M$\lambda$ }} \\
\multicolumn{13}{@{}l@{}}{\footnotesize{$^\mathrm{e}$ Recording rate: 256\,Mbit\,s$^{-1}$, 2 bits per sample}} \\
\multicolumn{13}{@{}l@{}}{\footnotesize{$^\mathrm{f}$ Dedicated full-track experiment on M\,87 (see  \citealt{kovalev07}); the array included a single VLA antenna (Y1) }}\\
\multicolumn{13}{@{}l@{}}{\footnotesize{$^\mathrm{g}$ Recording rate: 128\,Mbit\,s$^{-1}$, 1 bit per sample}} \\
\multicolumn{13}{@{}l@{}}{\footnotesize{$^\mathrm{h}$ Single polarization hand (LL) observations only}} \\
\multicolumn{13}{@{}l@{}}{\footnotesize{$^\mathrm{i}$ Recording rate: 512 Mbit\,s$^{-1}$, 2 bits per sample}} \\
\multicolumn{13}{@{}l@{}}{\footnotesize{$^\mathrm{j}$ Antenna NL missing}} \\
\multicolumn{13}{@{}l@{}}{\footnotesize{$^\mathrm{k}$ Antenna SC missing}} \\
\multicolumn{13}{@{}l@{}}{\footnotesize{$^\mathrm{l}$ Antenna HN missing}} \\
\end{tabular*}
\label{tab:epoch_list}
\end{table*}

One of the most studied AGN is M\,87 (also known as Virgo\,A), a nearby elliptical galaxy 
located in the Virgo cluster. It hosts a very powerful one-sided 
jet emerging from the central region, and was also the first extragalactic jet to be discovered \citep{curtis18}. The synchrotron-emitting nature of the M\,87 jet was suggested by \citet{baade56}. Observations show that M\,87 
contains $2.4\times10^{9}$ $M_{\odot}$ within a 
0.25$^{\prime\prime}$ radius, which suggests the presence of a SMBH in its 
nucleus \citep{harms94}. Due to its proximity (16.4\,Mpc, z=0.00436, 
1\,mas\,=\,0.08\,pc, 1\,mas\,yr$^{-1}$\,=\,0.26\,$c$, \citealt{jordan05}), 
M\,87 is an ideal candidate for studying AGN phenomena, and has been 
monitored at many different wavelengths  over the last several decades. 
In the observed one-sided jet of M\,87 \citep{shklovsky64}, superluminal motion was reported from {\it Hubble Space Telescope} 
({\it HST\/}) observations within the innermost 6$^{\prime\prime}$ of the jet 
with apparent speeds of 4\,$c$ to 6\,$c$ \citep{biretta99}. Discrepant 
speeds were reported from VLBA observations at 7\,mm with values between 0.25\,$c$ to 0.4\,$c$ 
\citep{ly07}, and a value of 2\,$c$ or even larger \citep{acciari09}. VLBA $\lambda$2\,cm observations showed apparent speeds $<$\,0.05\,$c$ from 1994 to 2007
\citep{kovalev07,lister09b}. Therefore, the kinematical 
properties of the jet in M\,87 remain an important topic of discussion.

In 1999 \textit{HST} observations revealed a bright knot in the jet located 1$^{\prime\prime}$ (projected distance of 0.08\,kpc) away from the core.
This feature, named HST-1, is  active in the radio, 
optical and X-ray regimes. VLBA $\lambda$20 cm observations showed that HST-1 has sub-structure and appears to contain superluminal components moving at speeds up to 4\,$c$ \citep{cheung07}. These observations suggest that HST-1 is a collimated shock in the AGN jet. Furthermore, recent multi-wavelength observations were used to show that HST-1 could 
be related to the origin of the TeV emission observed in M\,87 in 2005 by the HESS telescope \citep{aharonian06}. 
Comparing data taken in the near ultraviolet by \citet{madrid09}, soft X-rays (\textit{Chandra}) and 
VLA $\lambda$2\,cm observations \citep{cheung07, harris06}, the 
light curves of HST-1 reached a maximum in 2005, while the resolved 
core showed no correlation with the TeV flare. Therefore, the TeV 
emission from M\,87 was suggested to originate in HST-1 \citep{harris08}. Based on those findings, \citet{harris08} suggest that HST-1 has a blazar nature. However, \citet{acciari09} reported rapid TeV flares from M\,87 in February 2008, which may have originated in the core, instead of HST-1, which remained in a low state during the flares in 2008. Recently, the \textit{Fermi} Large Area Telescope (LAT) team reported the detection of M\,87 at gamma-ray energies \citep{abdo09}. 

The AGN standard model considers the blazar behavior to originate 
at the vicinity of the SMBH. However, HST-1 is 80\,pc away from 
the core. If the HST-1 blazar hypothesis is true, this would pose a challenge to current AGN models. In this paper, we examine 
this hypothesis with $\lambda$2\,cm VLBI wide-field imaging of the 
HST-1 feature.  We present the description of our VLBI 
observations and the corresponding data reduction in Sect. \ref{sec:obsana}; the results are presented in Sect. \ref{sec:results}; the  discussion is presented
in Sect. \ref{sec:discussion}.  Finally, a short summary is 
presented in Sect. \ref{sec:summary}.  Throughout this paper, we use the term ``core'' as the apparent origin of AGN jets that commonly appears as the brightest feature in VLBI radio images of blazars \citep{lobanov98,marscher08}. We use a cosmology with $\Omega_{m}=0.27, \Omega_{\Lambda}=0.73$, and $H_{0}=71 \mathrm{km}\,\mathrm{s}^{-1}$\,Mpc$^{-1}$ \citep{komatsu09}.

\section{VLBI Observations and Data Analysis \label{sec:obsana}}

M\,87 has been monitored at $\lambda$2\,cm with the VLBA since 1994 
by the 2\,cm Survey/MOJAVE programs\footnote{\tt http://www.physics.purdue.edu/MOJAVE/} \citep{kellermann04, 
lister09a}. We re-analyzed 12 epochs 
of these monitoring program data obtained
after late 2001, together with three observing sets of targeted 
observations on M\,87 in 2000 (see Table \ref{tab:epoch_list}). The 2\,cm Survey/MOJAVE epochs from 2001 to 2009 were snapshot observations with a total integration time from 20 minutes to one hour, and were broken up into about six-minute-long scans in order to have a better $(u,v)$ coverage; the three epochs in 2000 were full-track observations with eight-hour integration time and included a single VLA antenna (Y1).

The VLBA data were processed by following the standard procedures in AIPS cookbook, as described in detail by \citet{kovalev07, lister09a}. The data were fringe-fitted before the imaging process. The data from MOJAVE project epochs after 2007.61 were processed using the pulse calibration signals to align the phases instead of fringe fitting, because the positions of the sources and the VLBA antennas are well determined \citep{petrov08, petrov09} , and the station clocks are known to be stable. However, for our wide-field imaging purposes, fringe fitting was necessary for a better determination of the phase and rate across the observed band, and therefore to better image the extended jet of M\,87. Moreover, it ensures the homogeneity of all the datasets.

In order to accurately image HST-1 with VLBI, there are two issues that need to be considered. First, 
HST-1 lies $\sim$800 $\lambda$2\,cm beamwidth away from the brightest feature (the VLBA 
core). For this reason, time or frequency data averaging would produce time and 
bandwidth smearing in the HST-1 region. Second, based on earlier 
VLBA $\lambda$20\,cm and VLA $\lambda$2\,cm observations of HST-1 
\citep{cheung07}, we expect the total 2\,cm flux density to be at the milliJansky level, which is much weaker than the total flux 
density of the inner jet ($\sim$2.5\,Jy). To detect HST-1, 
we need to image the inner jet region with its extended structure 
(over tens of milliarcseconds); otherwise, the sidelobes from the 
core would cover the HST-1 emission. To reach this goal, we applied 
natural weighting and $(u,v)$-tapering on the whole dataset.

\begin{figure*}[!Ht]
 \centering
 \begin{minipage}[t]{.49\textwidth}
   \centering
   \includegraphics[width=1\textwidth]{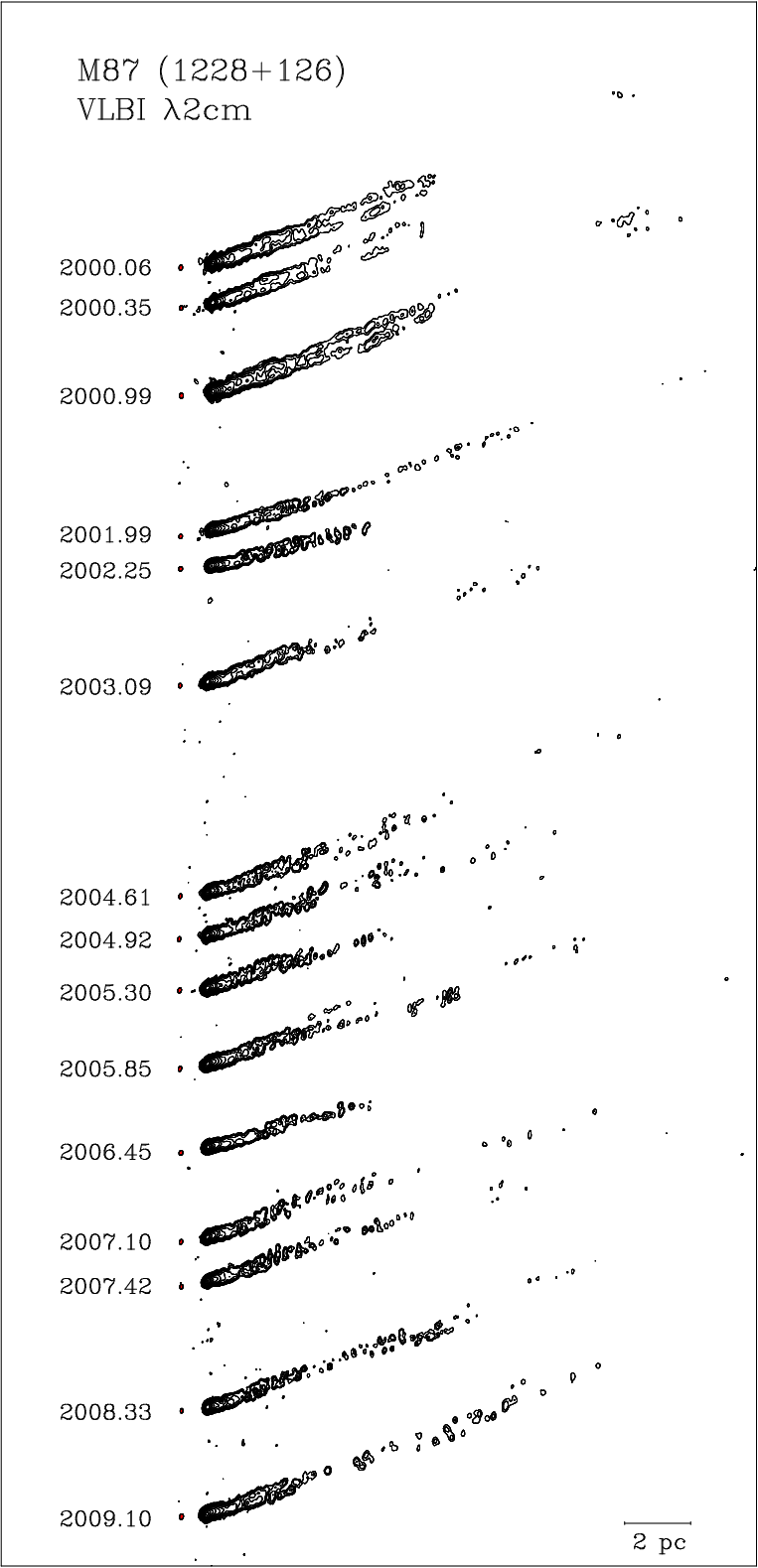}
   \caption{Images of the central 220\,mas of the
M\,87 jet (beam A, resulting from a $(u,v)$-taper of Gaussian 0.3 at 200\,M$\lambda$ and natural weighting), spaced by their relative time intervals. The images are plotted with the same size scale. The contour levels
all increase by factors of $\sqrt2$, and the lowest contours are (from 2000.06
to 2008.33): 0.5, 0.7, 0.5, 1.1, 1.3, 0.9, 0.9, 0.9, 0.9, 0.9, 1.1, 1.0, 1.2,
0.9, 0.9\,mJy\,beam$^{\mathrm{-1}}$, respectively. The restoring beams and brightness peak values are given in Table\ref{tab:epoch_list}.}
   \label{fig:field1}
 \end{minipage}
 \hspace{0.1cm}
 \begin{minipage}[t]{.49\textwidth}
   \centering
   \includegraphics[width=1\textwidth]{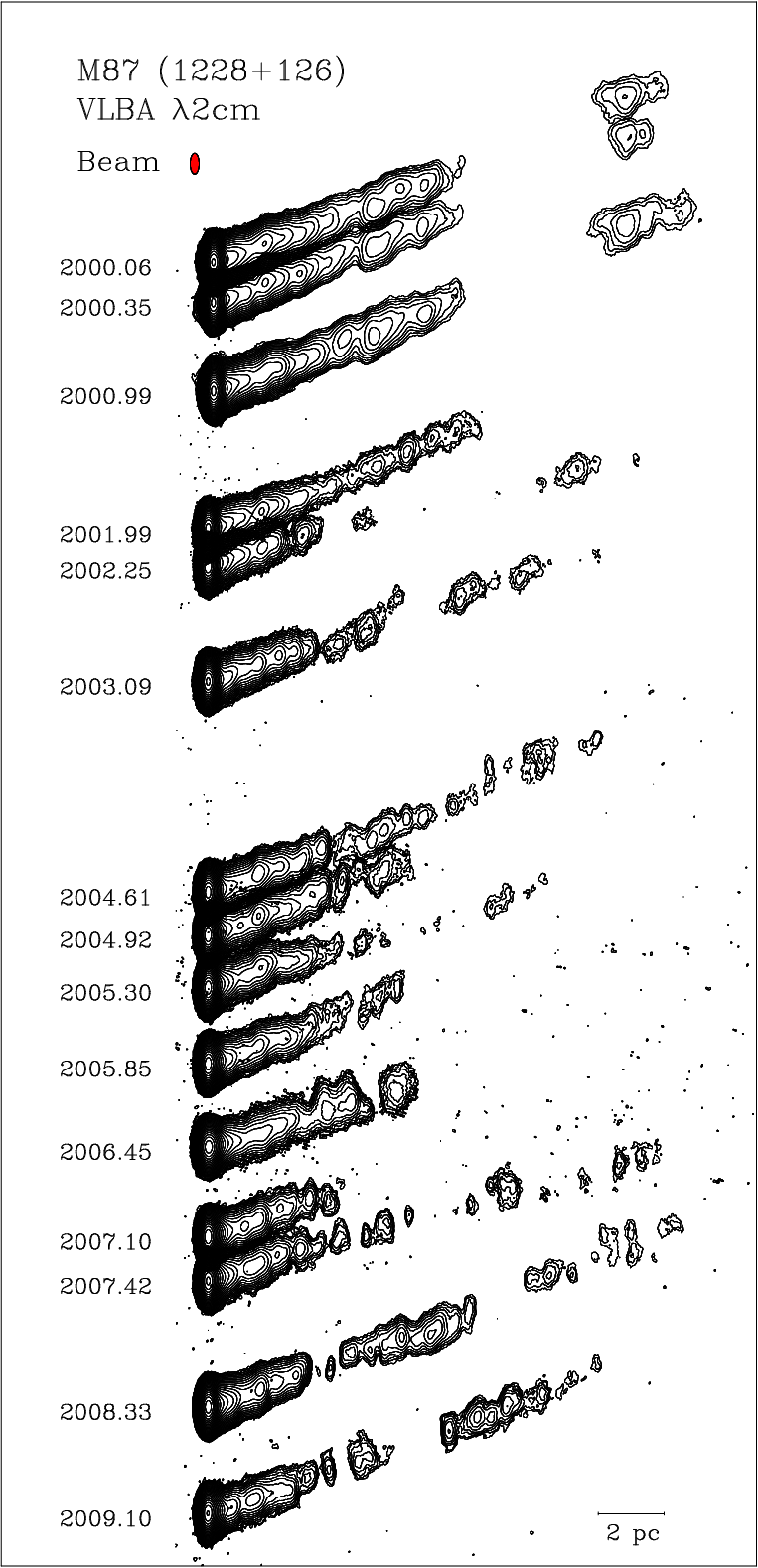}
   \caption{Downgraded resolution images of the central 220\,mas of the
M\,87 jet with beam B (8$\times$3.4\,mas), spaced by their relative time
intervals. The contour levels all increase by factors of $\sqrt2$, and the
lowest contours are (from 2000.06 to 2008.33): 0.5, 0.7, 0.5, 1.1, 1.3, 0.9,
0.9, 0.9, 0.9, 0.9, 1.1, 1.0, 1.2, 0.9, 0.9\,mJy\,beam$^{\mathrm{-1}}$, respectively.}
   \label{fig:field1_bbeam}
 \end{minipage}
\end{figure*}

First, we averaged the data 16-sec in time and averaged all channels in each IF in frequency, and obtained the inner jet initial model by applying phase and amplitude self-calibration using DIFMAP. The time-smearing effect of 16 seconds begins to be relevant at a distance of 200\,mas from the field center, which is beyond the inner jet scale, and did not affect our data analysis at this stage. When the inner jet model was reasonably good, we applied the obtained CLEAN-components to the un-averaged data for the first amplitude-and-phase self-calibration using AIPS. Natural weighting and tapering with a Gaussian factor of 0.3 at a radius of 200 M$\lambda$ in the $(u,v)$-plane was applied to the whole dataset, and the resultant beam sizes are shown in Table \ref{tab:epoch_list}. If we remove tapering and apply uniform weighting, the inner jet is better resolved but less flux density is recovered. In this case, the image noise level is too high to detect the HST-1 feature. 

To better identify the extended component, and since full-resolution, uniform-weighting imaging did not produce detections of this feature, we chose the approach of tapering and natural weighting. Furthermore, we downgraded the resolution to a larger beam size of 8$\times$3.4 mas in P.A. 0$^{\circ}$ This beam size was chosen specifically for comparison with the VLBA $\lambda$20\,cm observations of \citep{cheung07} In the following Sections,  we label the smaller beams  as \textbf{beam A}, noting that the beam sizes are slightly different in each epoch. These are listed in Table \ref{tab:epoch_list}. \textbf{Beam B} refers to the beam size of 8$\times$3.4\,mas.

\begin{figure}[Ht]
\centering
\resizebox{\hsize}{!}{\includegraphics{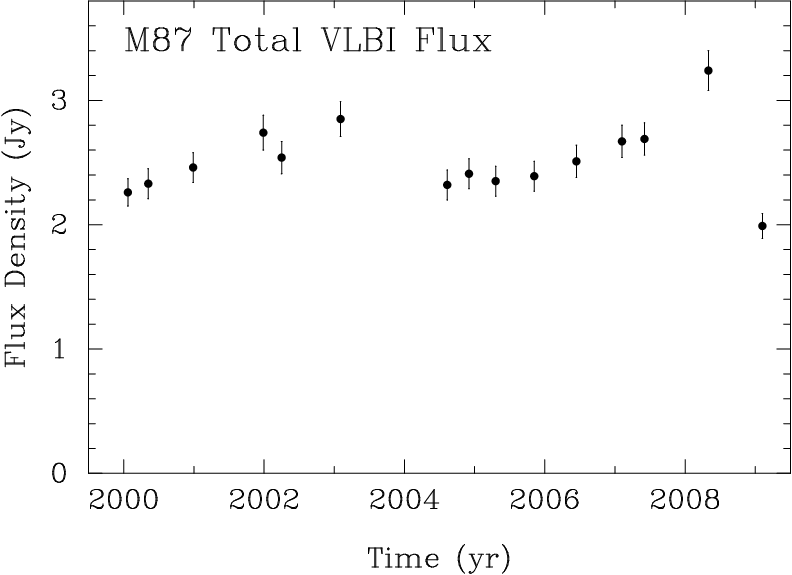}}
\caption{Total VLBA flux density of M87 at $\lambda$2\,cm versus time.}
\label{fig:jet_total}
\end{figure}

\section{Results \label{sec:results}} 
We completed the wide-field imaging of 15 epochs of M\,87 VLBA $\lambda$2\,cm obtained between 2000 and 2009. Each epoch was imaged using beam A and beam B (see Sect. \ref{sec:obsana}), and for each beam size, we applied two image cleaning fields, one for the M\,87 inner jet region, and another for the HST-1 region, which was phase-shifted as $-$788.5\,mas in right ascension and 348.9\,mas in declination. We performed deep cleaning iterations until the image rms reached the expected thermal noise value. The rms values of the final images are listed in Table \ref{tab:epoch_list}. Three epochs in the year 2000 are full-track observations with an on-source time of 476 minutes, and have the lowest rms.

\subsection{Imaging of the inner jet of M\,87}
With tapering and natural
weighting of the data, the extended inner jet was imaged over a region 100-200 mas in size, which varied according to the different sensitivity of each epoch. The inner jet images are shown in Fig.\,\ref{fig:field1} (beam A) and Fig.\,\ref{fig:field1_bbeam} (beam B). The inner jet structure has an average total flux density of $\sim$2.5\,Jy, with overall flux density changes up to 1\,Jy during 2000 to 2009, as shown in Fig.\,\ref{fig:jet_total}. We estimate the error in flux densities to be 5\%, based on the typical amplitude calibration accuracy of the VLBA (\citealt{kovalev05} and \citealt{VLBA_summ09}). In February 2008, multiple very high energy flares were detected by the multi-wavelength campaign of HESS, MAGIC, VERITAS; meanwhile, VLBA $\lambda$7\,mm observations detected a flare in the VLBA core simultaneously \citep{acciari09}. We have also seen the 2008 flare from the core in our VLBA $\lambda$2\,cm data, and therefore confirmed this result (see Fig.\,\ref{fig:jet_total}). 

The inner jet showed structural changes during the period of our observations. However, the study of the inner jet is beyond the scope of this paper (see e.g., \citealt{kovalev07}, \citealt{walker08}, and \citealt{lister09b} for a detailed discussion of the structure and kinematics of the inner jet).

\subsection{Imaging of the HST-1 region}
\begin{figure}[Ht]
 \resizebox{\hsize}{!}{\includegraphics{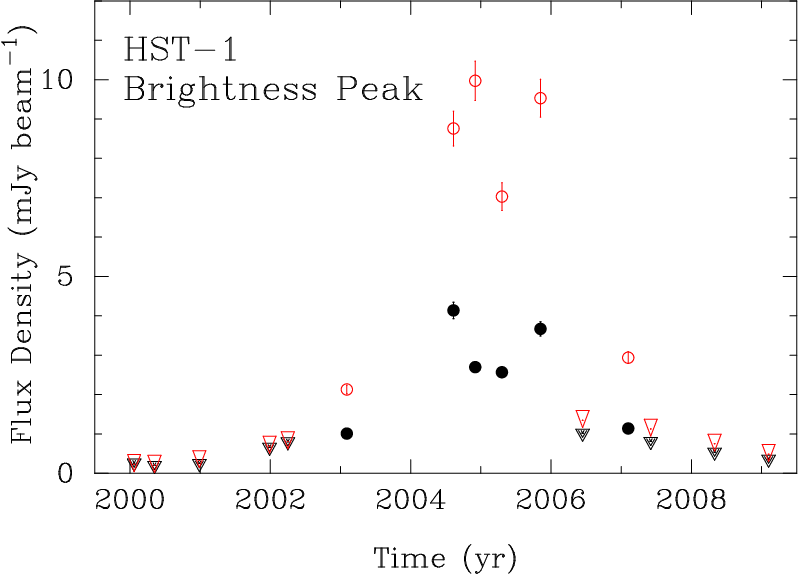}}
 \caption{Brightness peak value at VLBI $\lambda$2\,cm of the HST-1 region versus time. The values for epochs with no confident detection (inverted triangles) are upper limits, whereas the epochs with a detection are marked as dots. The results of using beam A are in black closed circles/triangles, while the results of using beam B are in red open circles/triangles. }
 \label{fig:hst1_peak}
\end{figure}

HST-1 was detected in six of the 15 epochs we analyzed from 2003 to early 2007 (detection limit: 5$\sigma$). As discussed in Sect. \ref{sec:discussion}, this feature was too faint in most epochs, and the image noise was too high to allow us to to detect it. Fig.\,\ref{fig:hst1_2beam} shows hybrid maps of this region. The HST-1 images produced using beam A show that the brightest component has a size of $\sim$10\,mas down to the lowest contours, while the images produced using beam B reveal an extended structure $\sim$50\,mas in size, which is comparable with VLBA $\lambda$20\,cm observations \citep{cheung07}. The feature has a total flux density that varies between 4 and 24\,mJy (Fig.\,\ref{fig:hst1_total}). The peak surface brightness varies from 1 to 4 mJy\,beam$^{-1}$ (beam A) and 2 to 10 mJy\,beam$^{-1}$ (beam B), as shown in Fig.\,\ref{fig:hst1_peak}. Epochs which have no detection are marked as upper limits (inverted triangles) in Figs.\,\ref{fig:hst1_total} and \ref{fig:hst1_peak}.

In 2003.09, HST-1 is marginally detected with both beam sizes; the brightness peak is $\sim$1\,mJy\,beam$^{\mathrm{-1}}$, and the total flux density is 3.7\,mJy. During 2004.61--2005.85, 4 epochs show significant detections of HST-1 with beam A and beam B, where the peak component remains dominant, and the total flux density fluctuation is $\sim$30\%. In 2007, the structure of HST-1 becomes more extended and complex, reaching a total flux density of 10\,mJy and developing two distinct peaks (peak \textbf{a} and peak \textbf{b} in Fig.\,\ref{fig:hst1_2beam}). 

For the epochs with no detections, we derived upper limit on the total VLBI flux density of HST-1. These were estimated based on the rms noise values of each epoch. The maximum amount of flux density hidden below the noise would be the rms value times the size of the HST-1 emission region (in units of beam size). Based on our HST-1 detection, we assume that the size of the HST-1 emission region is about 20 times the beam size, and we derived the upper limit on each epoch accordingly. We also derived the upper limit of the brightness temperature in this region as 9$\times10^{6}$ K at $\lambda$2\,cm.



\subsection{Spectral properties of HST-1}

\begin{figure*}[!Htp]
\centering
\includegraphics[width=0.82\textwidth]{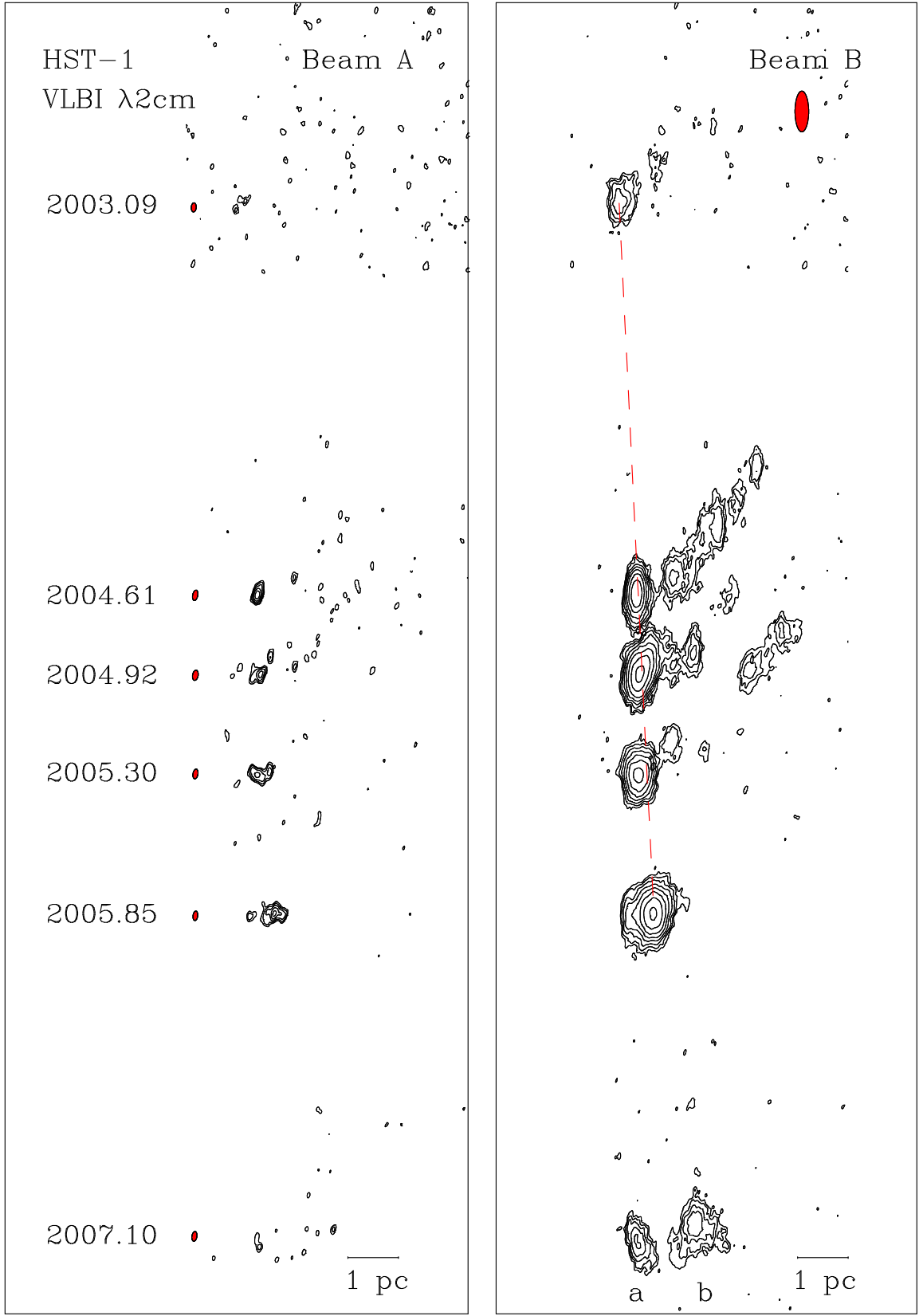}
\setlength{\floatsep}{50pt}
\caption{Images of the HST-
1 region restored with beam A (left panel) and beam B (right panel). The distance between epochs is proportional to the relative time interval, and the images are plotted with the same size scale. The contour levels increase by successive factors of $\sqrt2$, and the lowest contours for both images are (2003.09 to 2007.10): 0.6, 0.8, 0.8, 0.8, 0.8, 0.7 mJy\,beam$^{-1}$, respectively. The beam sizes of the beam A images are plotted to the right of each epoch label, while the beam size of the beam B images is shown in the upper-right corner. The HST-1 field is phase-shifted $-$788.5\,mas in right ascension (RA) and 348.9\,mas in declination (DEC). These images were obtained after performing self-calibration in the imaging process. In the right panel, \textbf{a} and \textbf{b} represent the two peaks of HST-1 in epoch 2007.10. The red dashed lines illustrates the linear fit yielding HST-1's projected apparent speed (see Fig. \ref{fig:hst1_motion}).}
\label{fig:hst1_2beam}
\end{figure*}

\begin{figure*}[Ht]
 \centering 
 \begin{minipage}{.48\textwidth}
   \centering
   \includegraphics[angle=270, width=1\textwidth]{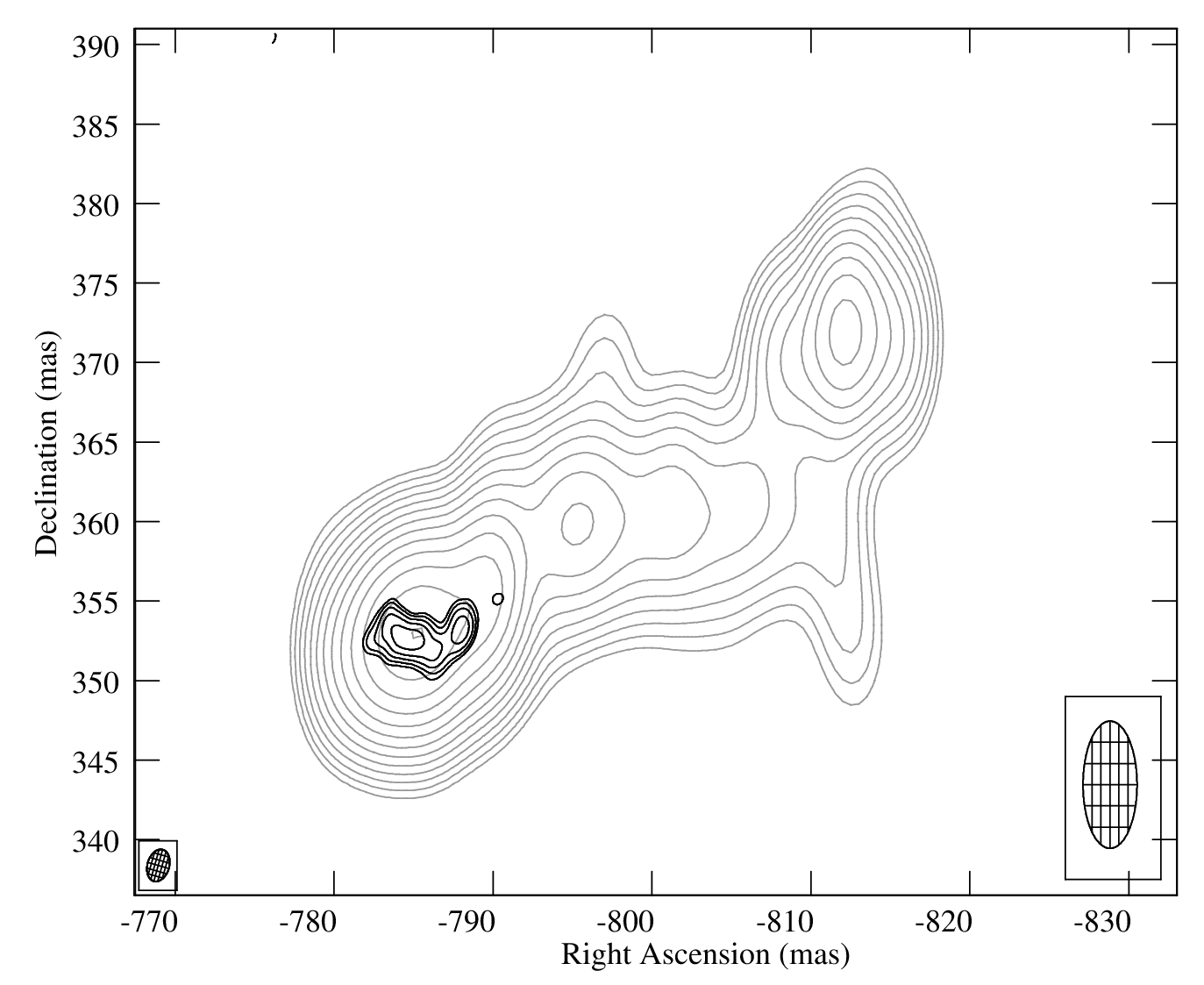}
 \end{minipage}
 \hspace{0.05cm}
 \begin{minipage}{.48\textwidth}
  \centering
  \includegraphics[angle=270, width=1\textwidth]{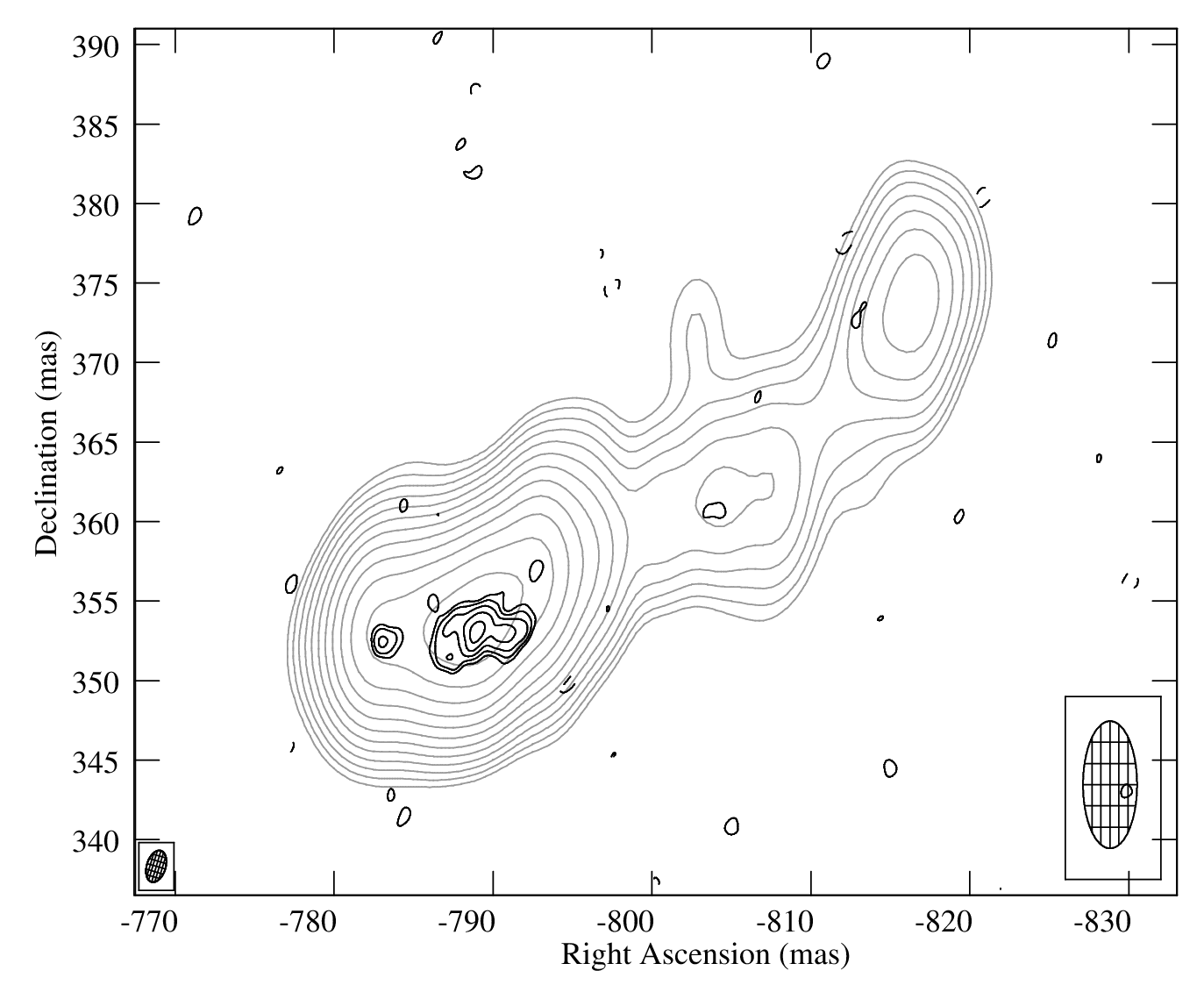}
 \end{minipage}
 \begin{minipage}{1\textwidth}
   \caption{VLBA images of the HST-1 region in M\,87 at $\lambda$\,2\,cm (this paper: black contour, beam size at half power level\,=\,2$\times$1\,mas, P.A.\,=\,$-16^{\circ}$, plotted bottom left) and $\lambda$\,20\,cm (\citet{cheung07}: gray contour, beam size\,=\,8$\times$3.4\,mas, P.A.\,=\,0$^\circ$, plotted bottom right). Left-hand panel: epoch 2005.30 ($\lambda$\,2\,cm, peak surface brightness: 3.4\,mJy\,beam$^{-1}$) and 2005.35 ($\lambda$\,20\,cm, peak: 45\,mJy\,beam$^{-1}$); right-hand panel: epoch 2005.85 ($\lambda$\,2\,cm, peak: 3.3\,mJy\,beam$^{-1}$) and 2005.82 ($\lambda$\,20\,cm, peak: 42\,mJy\,beam$^{-1}$). The lowest contour is 0.7\,mJy\,beam$^{-1}$, and the contour levels are separated by a factor of $\sqrt2$. The images of two frequencies were registered using the peak of the inner jet of HST-1 as a common reference point. }
 \label{fig:hst1_overlay}
 \end{minipage}

\end{figure*}

Figure\,\ref{fig:hst1_overlay} illustrates two overlaid images of HST-1 at $\lambda$20\,cm (\citealt{cheung07} and Cheung, priv.\ comm.) and $\lambda$2\,cm (beam A) of two adjacent epochs. The images were co-aligned using the peak of the inner jet as a common reference point. This figure shows that our observations are resolving the HST-1 peak component. We use those epochs to derive the HST-1 spectral properties by producing a rendition of the $\lambda$2\,cm image using the same restoring beam as the $\lambda$20\,cm image, namely, 8$\times$3.4\,mas at a position angle of 0$^{\circ}$ (beam B; Fig.\,\ref{fig:hst1_2beam}, right panel). Table \ref{tab:spectral_index} lists the epochs of $\lambda$20\,cm and $\lambda$2\,cm observations and the corresponding spectral index $\alpha$, where S$_{\nu}$\,$\propto$\,$\nu^{+\alpha}$. The uncertainties of the spectral indices are formal errors, and are estimated using standard error propagation methods. When interpreting these values, one has to keep in mind that the resolution of $\lambda$20\,cm and $\lambda$2\,cm are very different. First, the frequency is different by one magnitude, which results in small formal error bars of the spectral index. Second, the incomplete $(u, v)$ coverage on short VLBA baselines might cause a missing flux density in the HST-1 region at $\lambda$2\,cm (see Sect. \ref{sec:spectra} for a discussion).  The average value of the derived spectral index is $\alpha \sim$ $-$0.78.

\begin{table}[h]
\centering
\caption{HST-1 $\lambda$20$-$$\lambda$2\,cm spectral index$^{\mathrm{a}}$ }
\begin{tabular*}{0.4\textwidth}{@{}ccccc@{}}
\hline
\hline
\multicolumn{2}{c}{Epoch}        &  \multicolumn{2}{c}{S$_{\mathrm{\nu}}$ \footnotesize{[mJy]}} & $\alpha$  \\
$\lambda$2\,cm & $\lambda$20\,cm &  $\lambda$2\,cm & $\lambda$20\,cm &           \\
\hline
2005.30        & 2005.35         &  19.7$\pm$1.0             &   111$\pm$6$^{\mathrm{b}}$  &  $-$0.75$\pm$0.03 \\
2005.85        & 2005.82         &  19.9$\pm$1.0             &   126$\pm$6$^{\mathrm{b}}$  &  $-$0.80$\pm$0.02 \\
\hline
& & & & \\
\multicolumn{5}{@{}l@{}}{\footnotesize{$^\mathrm{a}$ The spectral index and its errors are formal, see section \ref{sec:spectra}}  } \\
\multicolumn{5}{@{}l@{}}{\footnotesize{$^\mathrm{b}$ 5\% error assumed for VLBI observations }} \\
\end{tabular*}
\label{tab:spectral_index}
\end{table}

\subsection{HST-1 kinematics}
To study the kinematics of HST-1, we fitted the peak of HST-1 and the M\,87 core in the image plane with a Gaussian component using IMFIT in AIPS. As shown in Fig.\,\ref{fig:hst1_2beam}, it is difficult to identify moving components in this region, because the detections are weak. Therefore it is difficult to derive an accurate apparent speed for the HST-1 subcomponents. For this reason, we use the relative position between the peak of HST-1 and the M\,87 core to estimate the apparent speed of HST-1. Moreover, among the 6 epochs of detection, HST-1 was almost resolved out in epoch 2007.10, and appeared to have a double-peak morphology in beam B image (Fig.\,\ref{fig:hst1_2beam}). By considering that the HST-1 detection in 2007.10 was weak, and the double-peak structure might be due to the effect of convolution, we excluded this epoch from the kinematic analysis (see Sect. \ref{sec:hst1_speed_discussion} for a discussion). As shown in Fig. \ref{fig:hst1_motion}, we plotted the peak positions of HST-1 with respect to the M\,87 core against time, and we estimated the position errors as the FWHM of the fitted Gaussian component. By applying a linear regression to the peak positions, we obtained a value for the projected apparent speed of the HST-1 feature of 2.36$\pm$1.19 mas\,yr$^{-1}$, which correspond to $\beta_{\mathrm{app}}$ = 0.61$\pm$0.31.




\begin{figure}[Ht]
 \resizebox{\hsize}{!}{\includegraphics{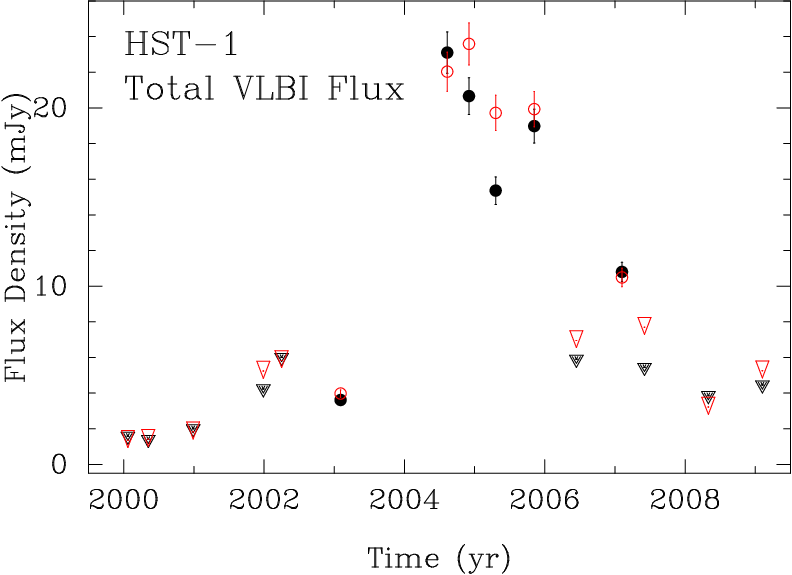}}
 \caption{Total VLBI $\lambda$2\,cm flux density of the HST-1 region versus time. The values for epochs with no confident detection (inverted triangles) are upper limits, whereas the epochs with a detection are marked as dots. The results of using beam A are in black closed circles/triangles, while the results of using beam B are in red open circles/triangles. }
 \label{fig:hst1_total}
\end{figure}

\section{Discussion \label{sec:discussion}}

\subsection{Variability timescale and HST-1 flaring region at parsec-scales}

We derived the variability timescale of the 2005 flare from HST-1 assuming that a single flare would produce a logarithmic rise and fall in the light curve. By fitting the available light curves from VLBI and VLA measurements before and after the maximum flux density during the flare, we derived corresponding variability timescales and the upper limit of the characteristic size where the flare was produced. We define the logarithmic variability timescale as $\tau_{\mathrm{var}} = dt/d[ln(S)]$, where $S$ is the flux density, $t$ is the time interval between observations in units of years \citep{burbidge74}. Next, we estimate the upper limit of the characteristic size of the emission region from light-travel time: $\theta_{\mathrm{char}}\,\approx\,0.13\,(1+\mathrm{z})\,\mathrm{D}^{-1}\tau_{\mathrm{var}}\,\delta$, where z is the redshift, D is the luminosity distance in Gpc, and $\delta$ is the Doppler factor \citep{marscher79}. Table\,\ref{tab:var_time} shows the result of the derived characteristic size of the 2005 flare in HST-1.  

\citet{harris03} estimated the Doppler factor $\delta$ of HST-1 to have a value between 2 and 5, based on decay timescales of \textit{Chandra} observations. In the same context, \citet{wang09} estimated the Doppler factor of HST-1 to be 3.57$\pm$0.51 by fitting the non-simultaneous spectral energy distribution of M87 using a synchrotron spectrum model. Simulations incorporating MHD models for the M\,87 jet have suggested that the recollimation shock formed close to the HST-1 position had a relatively low Doppler factor $\sim$1-2 \citep{gracia09}. If we use $\delta$=3.57, the derived characteristic sizes of HST-1 emission region during its flaring time are 20\,$<\theta_{\mathrm{rise}}<$\,31 and 54\,$<\theta_{\mathrm{fall}}<$\,100\,mas. The size scale of structural changes of HST-1 (see Fig.\,\ref{fig:hst1_2beam}) is between 20$-$50 mas (1\,pc = 12.5\,mas), which is within the derived characteristic size scale. However, if $\delta<$3.5, $\theta_{\mathrm{rise}}$ derived from VLBA $\lambda$2\,cm is smaller than 20\,mas, which would create a causality problem, since the largest structural change of HST-1 cannot be bigger than the upper limit of the information propagation time. Therefore, $\delta>$3.5 is needed based on the causality argument.

\begin{table}[Hb]
\centering
\caption{HST-1 variability timescale and characteristic size during the rise and fall of the 2005 flare.}
\begin{tabular*}{0.4\textwidth}{@{}lccll@{}}
\hline
\hline
Wavelength                   & $\tau_{\mathrm{rise}}$ & $\tau_{\mathrm{fall}}$ & $\theta_{\mathrm{rise}}$ & $\theta_{\mathrm{fall}}$  \\
                             &\scriptsize{[yr]}   & \scriptsize{[yr]} & \scriptsize{[mas]} & \scriptsize{[mas]}  \\
\hline
VLA $\lambda$2\,cm           & 1.5        & 4.9  & 8.7\,$\delta^{\mathrm{a}}$ & 28\,$\delta$     \\ 
VLBA $\lambda$2\,cm (beam A) & 1.0        & 4.2  & 5.6\,$\delta$ & 24\,$\delta$      \\ 
VLBA $\lambda$2\,cm (beam B) & 1.0        & 2.7  & 5.7\,$\delta$ & 15\,$\delta$      \\
\hline
 & & & & \\
\multicolumn{5}{@{}l@{}}{\footnotesize{$^\mathrm{a}$The characteristic size is related to the Doppler factor $\delta$. }} \\
\end{tabular*}
\label{tab:var_time}
\end{table}

If $\delta$ is less than 3.5 in the HST-1 region, we could explain the causality problem by the self-calibration procedure that we used while imaging HST-1. We applied self-calibration to the inner jet and HST-1 for all of the epochs with detections in beam A and beam B. Self-calibration works better with stronger objects, in our case, the marginal detection in epoch 2003.09 might not provide a satisfactory result in recovering extended emissions, comparing with stronger detections between 2004.61 and 2005.85, in which after self-calibration, we were able to recover more extended emission.


\begin{figure}[Hb]
 \resizebox{\hsize}{!}{ \includegraphics{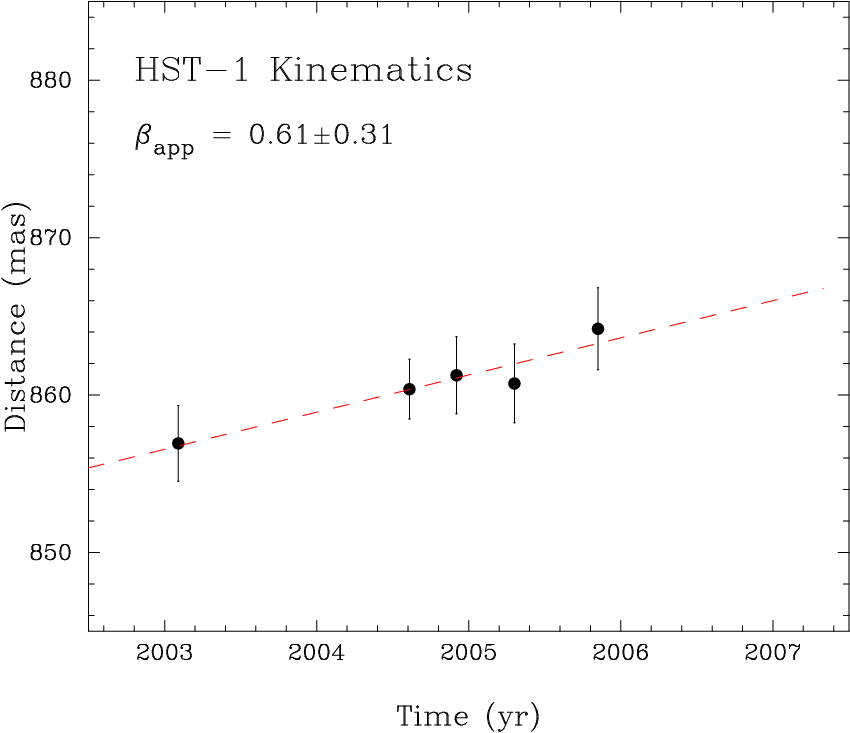}}
 \caption{The linear fit of HST-1 proper motion shown as the red dashed line. This plot illustrates the position of HST-1 component peaks with respect to the M\,87 core component as a function of time from 2003.09 to 2005.85. The fitted projected apparent speed of HST-1 $\beta_{\mathrm{app}}$ is shown at the upper-left corner of the plot.}
 \label{fig:hst1_motion}
\end{figure}

\subsection{Speeds of HST-1}
\label{sec:hst1_speed_discussion}
We used the peak position and fitted Gaussian errors to estimate the projected apparent speed of HST-1 $\beta_{\mathrm{app}}$ = 0.61$\pm$0.31 (see Fig. \ref{fig:hst1_motion}), which is sub-luminal. In the kinematic analysis, we excluded epoch 2007.10 for obtaining a robust kinematic result. However, to make our results more solid, we estimated the upper and lower limit of HST-1 apparent speed by including epoch 2007.10 as a test. The derived possible range of the apparent speed is 0.23 $<\beta_{\mathrm{app}}<$ 1.2, which is still consistent with a mildly relativistic jet motion.


To compare our kinematic results with previous findings, we show the positional evolution on the sky of the HST-1 peak at VLBA $\lambda$2\,cm with both beam A and beam B (see Fig. \ref{fig:hst1_2beam}), and components C1 and C2 at VLBA $\lambda$20\,cm (\citealt{cheung07} and priv. comm.). As illustrated, our derived apparent speed range and structural evolution in time at $\lambda$2\,cm are consistent with the $\lambda$20\,cm results. However, our sub-luminal speed measurements of HST-1 are inconsistent with the high apparent superluminal motions reported in \citet{cheung07} from VLBA $\lambda$20\,cm observations and with the \textit{HST} \citep{biretta99}. Nevertheless, in the VLBA $\lambda$20\,cm observations \citep{cheung07}, lower speeds were derived from some components in HST-1: c2 has $\beta_{\mathrm{app}}$\,=\,0.47\,$\pm$\,0.39, and HST-1d has $\beta_{\mathrm{app}}$\,=\,1.14\,$\pm$\,0.14. Our results are consistent with these findings.

\begin{figure*}[Ht]
 \centering
 \begin{minipage}[!b]{.48\textwidth}
   \centering
   \includegraphics[width=1\textwidth]{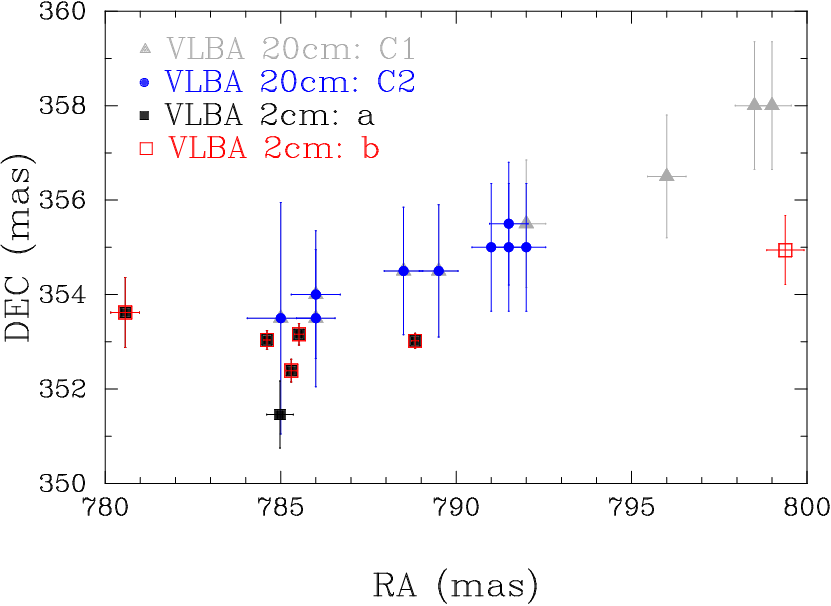}
 \end{minipage}
 \hspace{0.05cm}
 \begin{minipage}[!b]{.48\textwidth}
  \centering
  \includegraphics[width=1\textwidth]{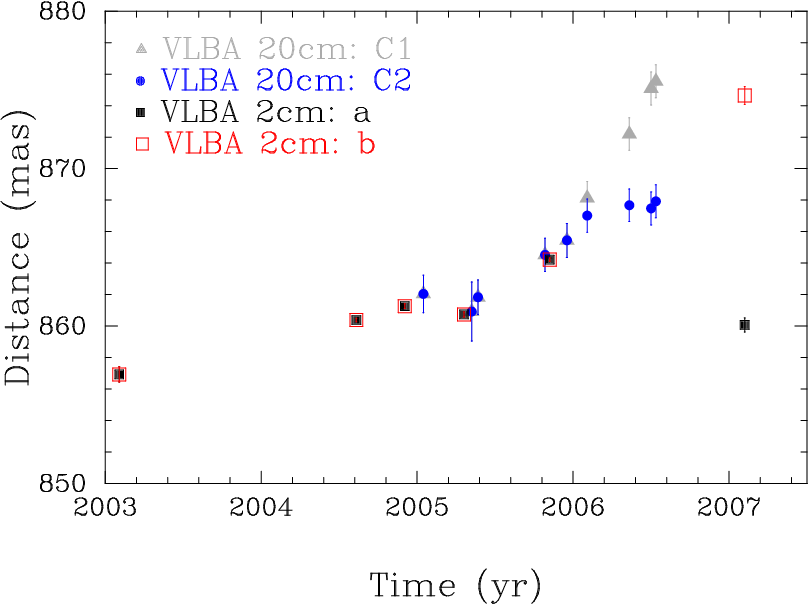}
 \end{minipage}
 \begin{minipage}[t]{1\textwidth}
   \caption{HST-1 sky positions (left) and radial distance as a function of time (right) relative to the M\,87 core. VLBA $\lambda$20\,cm (\citealt{cheung07} priv. comm.): time range of components C1 and C2 is from 2005.04 to 2006.53; VLBA $\lambda$2\,cm (beam B): time range of components \textbf{a} and \textbf{b} is from 2003.09 to 2007.10; components \textbf{a} and \textbf{b} are shown in Fig.\,\ref{fig:hst1_2beam}. }
 \label{fig:hst1_xyrt_position}
 \end{minipage}

\end{figure*}


\subsection{Detection limits}
HST-1 was not detected in epoch 2006.45. This epoch had only 22 minute integration time and the sampling rate was 128Mbit\,s$^{\mathrm{-1}}$. Therefore, the rms level (0.35 mJy\,beam$^{-1}$) was not low enough to detect HST-1, which was also fading according to our light curve (Fig.\,\ref{fig:hst1_total}).

The total flux density of HST-1 measured by the VLA at $\lambda$2\,cm reached its maximum value of $\sim$123\,mJy in 2005 \citep{harris06}. Our VLBA $\lambda$2\,cm results recovered a HST-1 total flux density of $\sim$23\,mJy, which is only 19\% of the VLA measurement. We conclude that the innermost region of HST-1 was resolved out with the long baselines of the VLBA. The low measured flux density suggests that HST-1 is very extended.


\subsection{HST-1 parsec-scale spectrum}
\label{sec:spectra}
We derived the spectral index of the HST-1 region based on VLBA $\lambda$2\,cm and $\lambda$20\,cm data (Table \ref{tab:spectral_index}). However, one should be cautious with these values. First, the two datasets have a frequency difference of a factor of 10, which results in small formal error bars of the derived spectral index. One should be aware that the spectral index here could provide us with a trend in the spectrum of HST-1 and the physical condition of the region, but not the absolute value. Second, we have found that the incomplete $(u,v)$ coverage on short VLBA baselines has resulted in missing flux in the HST-1 region at $\lambda$2\,cm. Therefore, the spectral index value of HST-1 region is a lower limit, and the results suggest that HST-1 was optically thin during the flare in 2005. If assuming HST-1 was a flat-spectrum source, there would have been $\sim$90 mJy missing flux in this region in our 2\,cm VLBA measurements.


\subsection{A blazar nature of HST-1}
The intensity of HST-1 reached a maximum in different wavebands in 2005, and the light curves from the VLA $\lambda$2\,cm, VLBA $\lambda$20\,cm, NUV, and X-ray observations all show the same tendency \citep{harris09, madrid09}. Our light curve of HST-1 at VLBA $\lambda$2\,cm (Fig.\,\ref{fig:hst1_total}) is consistent with the other observations. Although this trend could be taken as evidence that HST-1 is the source of the HESS-detected TeV flare in 2005 \citep{aharonian06}, its VLBA radio properties that we have derived do not support it. These include (i) a very low compactness of the dominant emission, (ii) low brightness temperature, (iii) sub-luminal motion, and (iv) possibly low optical depth across the feature at parsec-scales. Those are in contrast to the typical blazar core features, which tend to have flat or inverted spectral indices in cm-wave VLBA images. We observed that HST-1 has radio properties consistent with those commonly seen in jet components. TeV observatories cannot resolve the M\,87 core and HST-1 separately. To probe the origin of TeV emission, correlating variability between different bands is a powerful tool. Recent examples of this approach were successfully shown for a sample of blazars by \citet{kovalev09}, and for M\,87 by \citet{acciari09}.  Another approach is to apply physical models to multiwavelength observations to probe the TeV origin. There are models that suggest radio-TeV connections. For example, high energy flares could be generated from parsec-scale radio jets by inverse-Compton scattering of the photons and particles emitting from the core, and \citet{stawarz06} used this approach to explain the TeV flare from M87 in 2005. However, \citet{acciari09} reported about the VHE flare of M87 in 2008, and suggested that there might be correlations with the radio flare observed by VLBA at 43\,GHz from the core.  From our results, although we cannot fully discount, however, that HST-1 was the source of the TeV flare in 2005 based on our results alone, we do not favor the blazar nature for HST-1 suggested by \citet{harris08}.

\section{Summary \label{sec:summary}}
With our VLBA $\lambda$2\,cm data from 2000.06 to 2009.10, we have detected HST-1 during 2003.09--2007.10, which covered the multi-band flaring period of HST-1. The total flux density of HST-1 varied from 4-24 mJy in our detections; by comparing the images of VLBA $\lambda$2\,cm and $\lambda$20\,cm, we saw a steep spectrum with $\alpha>$\,$-$0.8 in this region. The projected apparent speed of HST-1 derived from the brightness peak position is 0.61$\pm$0.31$c$, which suggests a sub-luminal nature at VLBA $\lambda$2\,cm.  

Our results showed that HST-1 is extremely extended at parsec-scales, and has a steep spectrum. No compact feature with a brightness temperature higher than 9$\times10^{6}$ K is present in the $\lambda$2\,cm VLBA observations of this region of the M\,87 jet, which implies that HST-1 does not have the properties of a standard blazar core. Combining our findings, we do not find evidence of a blazar nature for HST-1 in the jet of M\,87.

\begin{acknowledgements}
We thank A.~Mor\'e, G.~Cim\`o, S.~M\"uhle, M.~A.~Garrett, R.~W.~Porcas, R.~C.~Walker, and C.~M.~Fromm 
for valuable comments and inspiring discussions. 
Special thanks are due to C.~C.~Cheung for providing the 
VLBA $\lambda$20\,cm images \citep{cheung07}, and D.~E.~Harris for providing VLA $\lambda$ 2\,cm light curve \citep{harris09}. We thank the anonymous referee for useful comments and suggestions.
This research was supported by the EU Framework 6 Marie Curie 
Early Stage Training program under contract number 
MEST/CT/2005/19669 ``ESTRELA''. C.~S.~Chang is a member of the International Max Planck Research School for Astronomy and Astrophysics.
Part of this project was done while Y.~Y.~K.\
was working as a research fellow of the Alexander von~Humboldt 
Foundation. Y.~Y.~K. was partly supported by the Russian Foundation for Basic 
Research (project 08-02-00545) and the Alexander von Humboldt return fellowship.
This research has made use of data from the 2\,cm\,Survey 
\citep{kellermann04} and MOJAVE programs that is maintained by the MOJAVE team \citep{lister09a}.
The MOJAVE project is supported under National Science Foundation grant
AST-0807860 and NASA \textit{Fermi} grant NNX08AV67G.
Part of this work made use of archived VLBA and VLA data obtained by 
K.~I.~Kellermann, J.~Biretta, F.~Owen, and W.~Junor.
The Very Long Baseline Array is operated by the USA National 
Radio Astronomy Observatory, which is a facility of the USA 
National Science Foundation operated under cooperative agreement 
by Associated Universities, Inc.
This research has made use of the NASA/IPAC Extragalactic Database (NED)
which is operated by the Jet Propulsion Laboratory, California Institute
of Technology, under contract with the National Aeronautics and Space
Administration. This research has made use of NASA's Astrophysics Data
System.
\end{acknowledgements}

\bibliographystyle{aa}
\bibliography{submitted}

\end{document}